\documentclass[aps,pra,preprint,groupedaddress,floatfix,nofootinbib]{revtex4-1}

\usepackage{graphicx,color,graphics}
\usepackage{dcolumn} 
\usepackage{amsmath}
\usepackage{amsfonts}
\usepackage{amssymb}
\usepackage{hyperref}
\usepackage{url}

\providecommand{\ignore}[1]{}

\newcommand{\epomean}{0.162}
  \newcommand{\epoerr}{0.008}\newcommand{\epoerrp}{(8)}
  \newcommand{\normepo}{0.108}\newcommand{\normepoerrp}{(5)} 
\newcommand{\epoAtoD}{0.144}\newcommand{\epoAtoDerrp}{(11)}
\newcommand{\epoCtoF}{0.185}\newcommand{\epoCtoFerrp}{(20)}
\newcommand{\epoDtoF}{0.237}\newcommand{\epoDtoFerrp}{(25)}

\newcommand{\epoprepmean}{0.086}\newcommand{\epopreperrp}{(22)}

\newcommand{\epgmean}{0.069}
  \newcommand{\epgerr}{0.017}\newcommand{\epgerrp}{(17)}
\newcommand{\epgAtoC}{0.048}\newcommand{\epgAtoCerrp}{(26)}
\newcommand{\epgAtoD}{0.071}\newcommand{\epgAtoDerrp}{(20)}
\newcommand{\epgCtoF}{0.120}\newcommand{\epgCtoFerrp}{(44)}
\newcommand{\epgDtoF}{0.090}\newcommand{\epgDtoFerrp}{(100)}

\newcommand{\epgprepmean}{0.132}\newcommand{\epgpreperrp}{(26)}

\newcommand{\epgoneleftmean}{0.010}\newcommand{\epgonelefterrp}{(2)}
\newcommand{\epgonerightmean}{0.007}\newcommand{\epgonerighterrp}{(2)}

\newcommand{\epgoneleftmeantwo}{0.009}\newcommand{\epgonelefterrptwo}{(2)}
\newcommand{\epgonerightmeantwo}{0.012}\newcommand{\epgonerighterrptwo}{(3)}

\newcommand{\siunit}[1]{{}\,\textrm{#1}}

\begin{document}

\title{Randomized Benchmarking of Multi-Qubit Gates}
\author{J. P. Gaebler}
\email[Electronic address: ]{john.gaebler@nist.gov}
\author{A. M. Meier}
\author{T. R. Tan}
\author{R. Bowler}
\author{Y. Lin}
\author{D. Hanneke}
\altaffiliation{Current address: Department of Physics, Amherst College, Amherst, MA, 01002-5000 USA}
\author{J. D. Jost}
\author{J. P. Home}
\altaffiliation{Current address: Institute for Quantum Electronics, ETH Zurich, 8093 Zurich, CH}

\author{E. Knill}
\author{D. Leibfried}
\author{D. J. Wineland}
\affiliation{National Institute of Standards and Technology, 325 Broadway, Boulder, CO 80305, USA}

\date{\today}

\begin{abstract}
  As experimental platforms for quantum information processing
  continue to mature, characterization of the quality of unitary gates
  that can be applied to their quantum bits (qubits) becomes
  essential.  Eventually, the quality must be sufficiently high to
  support arbitrarily long quantum computations.  Randomized
  benchmarking already provides a platform-independent method for
  assessing the quality of one-qubit rotations.  Here we describe an
  extension of this method to multi-qubit gates. We provide a
  platform-independent protocol for evaluating the performance of
  experimental Clifford unitaries, which form the basis of
  fault-tolerant quantum computing.  We implemented the benchmarking
  protocol with trapped-ion two-qubit phase gates and one-qubit gates
  and found an error per random two-qubit Clifford unitary of
  $\epomean \pm \epoerr$, thus setting the first benchmark for
  such unitaries.  By implementing a second set of sequences
  with an extra two-qubit phase gate at each step, we extracted an
  error per phase gate of $\epgmean \pm \epgerr$.  We conducted these
  experiments with movable, sympathetically cooled ions in a
  multi-zone Paul trap---a system that can in principle be scaled to
  larger numbers of ions.
\end{abstract}
\maketitle

\section{Introduction}
\label{sec:intro}

Quantum information processing (QIP) has the potential to solve
difficult problems in many-body quantum mechanics and mathematics that
lack efficient algorithms on classical computers.  However, achieving
useful QIP will require precise control of many qubits (two-level
quantum systems) and the ability to execute quantum gates (operations
that manipulate the quantum states of the qubits) with low error per
gate.  Here, the error per gate (EPG) is $\varepsilon = 1-F$, where
$F$ is the (average) gate fidelity defined as the uniform average over
pure input states of $\langle\psi|\rho|\psi\rangle$, where $\rho$ is
the (typically mixed) output state and $|\psi\rangle$ is the intended
output state (see Ref.~\cite{Dankert06}). A convincing demonstration of
the potential for practical fault-tolerant QIP should include
verification of consistent EPGs below a threshold of
$10^{-4}$~\cite{Preskill98,Knill10}.

So far, there has been substantial experimental progress on the basic
techniques needed for QIP, including the manipulation of small numbers
of qubits and the implementation of the basic quantum gates that are
needed to perform useful algorithms~\cite{Ladd10}. The main challenges
for QIP experiments that remain are to scale up to larger numbers of
qubits and to decrease the EPG below the fault-tolerant threshold.
Therefore, it is desirable to efficiently characterize or benchmark
the performance of multi-qubit QIP experiments so as to extract the
EPG of specific gates and enable comparison between different quantum
computing platforms.  With these goals in mind, we give a benchmarking
protocol for arbitrary numbers of qubits and show the results from an
experimental implementation for two qubits.  The protocol builds on
previous work that used randomized sequences of Clifford gates to
measure the EPG of one-qubit gates, first implemented in
Refs.~\cite{Knill08,Ryan09}.

Compared to techniques such as process
tomography~\cite{Chuang1997,Poyatos1996}, randomized benchmarking
offers several key advantages for characterizing EPGs of quantum
gates.  For example, while process tomography offers more complete
information about the performance of a gate, it does not scale
efficiently with the number of qubits in the system, it cannot readily
measure EPGs below the error probabilities of state preparation and
readout, and it does not verify performance of a gate in arbitrary
computational contexts.  In contrast, randomized benchmarking can
determine EPGs with a number of measurements that scales polynomially
with the number of qubits~\cite{Knill08,Magesan11}. Because randomized
benchmarking measures an exponential decay of fidelity as a function
of the number of gates in the sequences, errors in state preparation
and readout do not limit the minimum EPG that one can measure.  Also,
randomized benchmarking involves gates in the context of long
sequences of operations and therefore establishes an EPG that takes
into consideration a computational context similar to that expected in
the implementation of lengthy QIP algorithms. Because of these
advantages, randomized benchmarking following the protocols of
Refs.~\cite{Knill08,Ryan09} has been used to measure one-qubit gate
errors in a range of systems including trapped
ions~\cite{Knill08,Brown11}, superconducting
qubits~\cite{Chow08,Chow10}, liquid-state NMR~\cite{Ryan09}, and
neutral atoms in an optical lattice~\cite{Olmschenk10}.  Recently,
randomized benchmarking was used to measure an EPG of $2.0(2)\times
10^{-5}$ for one-qubit operations with a trapped ion~\cite{Brown11}.

A number of previous works have described properties of two-qubit
gates with various measurement techniques. With trapped ions, the
fidelity for creating a Bell state has been measured at
$0.83(1)$~\cite{Sackett2000}, $0.97(2)$~\cite{Leibfried03}, $0.89(3)$
~\cite{Haljan2005}, $0.83(3)$ ~\cite{Home2006} and
$0.993(1)$~\cite{Benhelm2008}.  Process tomography was used to look at
single and repeated applications of a two-qubit entangling
gate~\cite{Riebe2006,Home09}. The average fidelities were found to be
$0.938(3)$ for one and $0.882(4)$ for two gates in Ref.~\cite{Home09}.
Two-qubit gates have also been studied in other quantum computing
platforms including superconducting and photonic qubits (see
Ref.~\cite{Ladd10} and citations therein), with measured fidelities
ranging from $0.90$ to $0.99$, disregarding photon loss for photonic
qubits.  In a liquid-state NMR system, a randomized benchmarking
technique was used to study the errors of sequences of randomized
gates on three nuclear spins~\cite{Ryan09} and found EPGs of
$0.0047(3)$. The gates in this experiment were randomly chosen in a
platform-dependent way from a special-purpose probability distribution
where the probability of a two-qubit gate (the CNOT) was $1/3$.
However, gate sets vary by platform, and other experiments may choose
different probability distributions, for example to improve
randomization.  As a result the error probabilities from
Ref.~\cite{Ryan09} may be difficult to compare to those obtained in
future experiments.

The multi-qubit protocol we describe first establishes a
platform-independent error per operation (EPO)~\footnote{We use the
  convention that EPGs refer to processes that are intended to
  implement elementary quantum gates, whereas EPOs refer to processes
  that implement quantum circuits that may scale with the number of
  qubits. In both cases, the gates or circuits need to be specified to
  interpret reported values.} for Clifford unitaries by applying
random sequences of Clifford unitaries of varying lengths.  Here, a
Clifford unitary is any operator in the Clifford group defined below.
It then determines the EPG of individual gates of our choice by
inserting them into these sequences. The individual gates to be
characterized may depend on the platform. Of particular interest are
implementations of one of the standard universal two-qubit gates such
as the controlled-not (CNOT), phase gate or square-root of swap.  The
basic principles of the protocol are similar to the theoretical
randomized Clifford-based benchmarking sequences described and
analyzed in Ref.~\cite{Magesan11}. However, like the standard protocol
used in one-qubit benchmarking experiments so far, the last gate of
the sequence does not strictly reverse the effects of the previous
ones, thus enabling the protocol to detect certain large errors that
can otherwise masquerade as no errors.  In addition, we discuss the
practical aspects of choosing the random Clifford unitaries in the
sequence and extend the protocol to enable characterization of
specific gates, thus enabling diagnostics that were previously
unavailable in randomized benchmarking. The proposed protocol is
flexible without affecting the ability to compare results from
unrelated implementations.  We note that the theoretical relationship
between the protocol's EPGs and EPOs and the detailed physical noise
parameters is not known in general~\cite{Magesan11}. However, we
suggest that subject to simple consistency checks, the protocol's
reported EPGs and EPOs are nevertheless useful quantities for
comparison and reflect computationally relevant error behavior.

For our demonstration with trapped ions, we take advantage of a
multi-zone ion trap~\cite{Jost10}.  The universal two-qubit gate
chosen here is a phase gate, $\hat{G}$, implemented via a M\o lmer-S\o
rensen gate~\cite{Sorenson99} and acting as the diagonal matrix with
diagonal $[1,i,i,1]$ in the basis labeled by $|\uparrow
\uparrow\rangle$,$|\uparrow \downarrow\rangle$,$|\downarrow
\uparrow\rangle$,$|\downarrow \downarrow\rangle$, where
$|\downarrow\rangle$ and $|\uparrow\rangle$ represent the two
eigenstates of each qubit, with $\sigma_z|\uparrow\rangle=+1$,
etc. Qubit addressing is implemented by separation of the ions into
different wells, and long sequences of gates are supported by
sympathetic cooling techniques as required for the approach to
scalable ion-trap quantum computing described in
Refs.~\cite{Bible,Kielpinski02}.  The experiment extends the
technology demonstrated in~\cite{Home09} by using longer sequences of
gates and a different implementation of the phase gate~\cite{Lee05} to
act directly on a magnetic-field-insensitive transition in $^9$Be$^+$.

From sequences of up to seven Clifford unitaries, each requiring an
average of $1.5$ phase gates, we deduced an EPO of $\epomean\epoerrp$
for the Clifford unitaries and an EPG of $\epgmean\epgerrp$ for the
phase gates. Although we implemented relatively long sequences, the
experiment does not yet demonstrate stationary behavior because ion
loss prevented routine implementation of longer sequences.  There are
also indications that the errors increased with sequence length by two
to three standard deviations, with the EPOs ranging from
$\epoAtoD\epoAtoDerrp$ to $\epoCtoF\epoCtoFerrp$ and the EPGs from
$\epgAtoC\epgAtoCerrp$ to $\epgCtoF\epgCtoFerrp$ as the sequences
lengthened.  Our EPO sets the first benchmark for random two-qubit
Clifford unitaries. The EPG shows no improvement over the
gates used in~\cite{Home09}, but applies to gates used in
computationally relevant contexts in longer sequences.

The paper is structured as follows: We first describe the
protocol and its main features for two qubits. We then discuss the
experimental implementation of the protocol and show the experimental
results. The data-analysis methods are detailed next, followed by a
discussion of necessary consistency checks and estimates of physical
sources of error. We finally define the protocol for arbitrary numbers
of qubits, and make recommendations for how to apply and compare it
when qubit numbers vary.

\section{Benchmarking Protocol}
\label{sec:bench2}

Clifford unitaries are fundamental to most error-correcting procedures
envisioned for quantum computing (see, for example,
Ref.~\cite{Gottesman09}) and thus serve as a foundation on which
universal fault-tolerant quantum computing is built. As a result, a
large fraction of the fundamental processes in proposed quantum
computing architectures involve Clifford unitaries. The three main
features of the group of Clifford unitaries that make it useful for
our purposes are that its members have compact representations that
can be efficiently converted to circuits of elementary quantum gates,
outcomes of standard measurements of sequences of Clifford unitaries
can be efficiently predicted by classical computation, and the group
is sufficiently rich that error operators can be perfectly
depolarized. These features are explained in context below.

For a system of $n$ qubits, Clifford unitaries can be constructed by
combining one-qubit $\pm \frac{\pi}{2}$ rotations, defined as
$\hat{R}_{u}(\pm\pi/2)=e^{\mp i\frac{\pi}{4}\sigma_u}$ with $u=x,y$,
about the $\hat{x}$ and $\hat{y}$ axes, and two-qubit CNOT gates.
Alternatively, the Clifford unitaries are the members of the Clifford
group, which is defined as the set of unitaries $U$ with the property
that for every Pauli operator $P$, $UPU^\dagger$ is a signed product
of Pauli operators. We consider two gates or unitaries that differ
only by a global phase to be identical.

The randomized benchmarking protocol is an extension of ``Clifford
twirling''~\cite{Dankert06}. In the simplest instance of Clifford
twirling, an arbitrary quantum process $\cal{P}$ is sandwiched between
a random Clifford unitary $C$ picked from a uniform distribution and
its inverse $C^{\dagger}$. Alternatively, we can think of Clifford
twirling as averaging the process $C_i^\dagger \mathcal{P} C_i$ over
all elements $C_i$ in the set $\mathcal{C}$, of Clifford unitaries.
The key property of Clifford twirling is that this new process behaves
like one that uniformly depolarizes with some probability.  In other
words, a single parameter, the probability that a pure input state is
mapped to an orthogonal state, characterizes the new, average
process~\cite{Dankert06}. When $\cal{P}$ is a noisy implementation of
the identity gate, such as a long self-reversing sequence of gates, we
use this parameter as the definition of the average error of
$\cal{P}$. Clifford twirling can be generalized to learn the average
error of an arbitrary process $\cal{P}$ intended to implement a
specific Clifford unitary $U$: The inverting Clifford unitary
$C^{\dagger}$ that is applied after the process is modified to an
implementation of the unitary $UC^{\dagger}U^{\dagger}$. With this
modification, the net effect is $U$ if there are no errors in
$\cal{P}$. Because Clifford unitaries form a group, $U
C^\dagger U^\dagger$ is also Clifford.  The implementation of $U
C^\dagger U^\dagger$ should not rely on $\cal{P}$ to provide $U$, and it is
better not to decompose it into a composition of three processes
according to the given expression. This can be satisfied by first
evaluating the unitary operator $U C^\dagger U^\dagger$ as an element
of the Clifford group and then implementing it by an efficient
procedures for translating Clifford unitaries into quantum
circuits~\cite{Aaronson2004,Patel2008}. If $\cal{P}$ contains errors,
then the net process applies the unitary $U$ followed by a uniformly
depolarizing error whose parameter defines the average error of
$\cal{P}$.

Randomized benchmarking extends the idea of Clifford twirling from the
simple three-step sequence described above to randomized sequences of
Clifford unitaries with errors. These sequences consist of steps where
each step implements a randomly-chosen Clifford unitary and may have
errors.  Each step in the sequence simultaneously acts as a process
that undergoes twirling and contributes to the twirling of errors in
the other steps.  Under optimistic assumptions described later, each
step effectively behaves as an ideal unitary followed by a
depolarizing process.  The first goal is to establish the average
error per step, which can then be reported as the EPO for Clifford
unitaries.

The method for implementing the Clifford unitaries making up the steps
in the randomized sequences is up to the experimenter. Here we describe
our approach.  To improve stability of the twirling process and take
advantage of the typically lower errors of one-qubit gates, each step
is composed of a Clifford unitary preceded by a Pauli unitary, where
the two parts are chosen so that together they implement a uniformly
random Clifford unitary.  The choice of Pauli unitary on each qubit is
random and independent from the choice of Clifford unitary; each Pauli
unitary involves applying either no pulse or a major-axis $\pi$-pulse.
There are eight possible such pulses acting as $e^{\pm
  i\sigma_u\pi/2}$, where $\sigma_u$ is a Pauli matrix or the
identity: $\sigma_u\in\{\mathbf{1},\sigma_x,\sigma_y$,
$\sigma_z\}$. The sign in the exponent affects only the global phase
and results in two choices for each possible matrix in the exponent.
We keep the sign because in many cases including ours, the change in
sign can involve a physically different device setting, such as the
phase in a pulse generator that determines the orientation of the
fields that mediate the pulse. Each qubit's $\pi$ pulse is chosen
uniformly at random from the above eight pulses.

Because of this Pauli randomization procedure, it suffices to choose
the unitary in the second part of the step uniformly at random from
the Clifford group modulo the group of Pauli products.  For this we
can take advantage of the fact that the group of Pauli products is a
normal subgroup of the Clifford group, and the quotient group (of
Clifford unitaries modulo Pauli products) has a representation in
terms of binary symplectic matrices $M$ of dimension $2n\times 2n$
such that $MSM^T = S$ modulo $2$, where $S$ is a $2\times 2$ block
matrix with $n\times n$ blocks whose diagonal blocks are zero and
whose off-diagonal blocks are the identity, see, for example
Ref.~\cite{Gottesman1997,Nielsen2001}. The terminology is based on
Ref.~\cite{Calderbank1996}. For two qubits, there are $720$ such
matrices $M$.  Uniformly and randomly choosing from among these
matrices is computationally straightforward and efficient.

Determining an implementation of the Clifford unitary described by
such a matrix in terms of the elementary gates available in a
particular experiment is more challenging.  There are efficient
algorithms that translate an arbitrary symplectic binary matrix into
order of $n^2/\log(n)$ elementary one- and two-qubit
gates~\cite{Aaronson2004,Patel2008}, each of which can then be mapped
into experimentally available operations. However, there is strong
motivation to obtain shorter implementations, as this is a sure way to
improve the measured EPO.  While it is unlikely that optimal
implementations can be readily obtained for arbitrary numbers of
qubits, we used the following strategy for two qubits optimized for
our demonstration: By exhaustively listing compact circuits of
one-qubit Clifford gates and phase gates $\hat{G}$, we determined for
each of the $720$ symplectic binary matrices a circuit with the
minimum number of phase gates implementing the corresponding Clifford
unitary (modulo a Pauli product). On average, $1.5$ phase gates were
required.  These circuits were then translated into appropriate
actions in our ion-trap platform.

\begin{figure}[h]

\begin{center}
\begin{picture}(6,1.1)(0,0)
\put(0,.5){
\makebox(0,0)[l]{$
M_C = \left( \begin{array}{cccc}
	0 & 1 & 0 & 1 \\
	0 & 1 & 1 & 1 \\
	1 & 1 & 0 & 0 \\
	1 & 0 & 0 & 0 \\
\end{array} \right )$ }}
\put(1.85,0.5){
\makebox(0,0)[c]{\Large$\rightarrow$}}
\put(2.5,0.5){
\makebox(0,0)[l]{\includegraphics[height=1in]{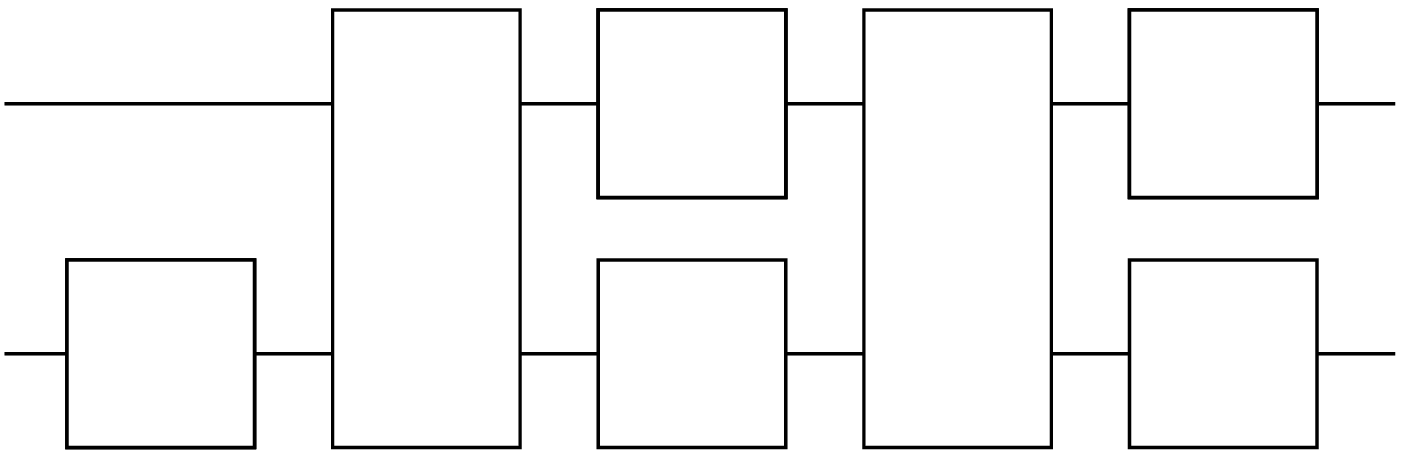}}}
\put(3.45,0.45){$\hat{G}$}
\put(4.63,0.45){$\hat{G}$}
\put(2.7,0.17){$\hat{R}_x(\frac{\pi}{2})$}
\put(3.88,0.72){$\hat{R}_x(\frac{\pi}{2})$}
\put(5.08,0.72){$\hat{R}_x(\frac{\pi}{2})$}
\put(3.9,0.17){$\hat{R}_y(\frac{\pi}{2})$}
\put(5.08,0.17){$\hat{R}_y(\frac{\pi}{2})$}
\end{picture}
\end{center}
\caption{\textbf{Example of Clifford unitary generation.}  First a
  random binary symplectic $4\times 4$ matrix $M_C$ is generated. In
  general, the size of such matrices is twice the number $n$ of
  qubits. They efficiently encode a Clifford unitary whose size is
  $2^n$ in general. The second step is to convert $M_C$ into a
  sequence of elementary gates that enacts the corresponding Clifford
  unitary and is suitable for implementation in the ion trap
  platform. We minimize the number of $\hat G =
  \textrm{diag}(1,i,i,1)$ gates in such sequences. The sequence found
  for the example $M_C$ shown is given on the right.  }
\end{figure}

Given the method for generating the random unitaries for one step, a
benchmarking experiment is configured by first deciding on a set of
lengths $l_1<\ldots<l_k$ that determine the numbers of steps in
sequences to be generated. The EPO is determined by fitting an
exponential decay to fidelities ( $1-E$, where $E$ is the error
probability) measured for each sequence length. The choice of lengths
therefore contributes to how well the EPO can be extracted. In
particular, there should be enough lengths for stable curve fitting,
and lengths much greater than the inverse EPO contribute little
additional information.  For each length $l$, many sequences of $l$
random steps are produced. At the end of each such sequence, a
randomized measurement step is added. This step consists of a Pauli
randomization followed by a Clifford unitary that inverts the $l$
preceding Clifford steps. The final Clifford unitary is chosen
independently of the Pauli randomization. This ensures that in the
absence of errors, the final state is again in the computational basis
but randomized. Which basis element it should be in can be computed by use of
standard efficient methods for simulating sequences of Clifford
unitaries~\cite{Gottesman1997}.  The sequence can then be
experimentally implemented after preparation of each qubit in the $-1$
eigenstate of $\sigma_z$ and followed by a measurement in the
$\sigma_z$ basis of each qubit.

One should implement sufficiently many runs of each sequence to have
good signal-to-noise on the inferred probabilities of getting a
correct or incorrect answer in the measurement for this particular
sequence. The process of generating and implementing random sequences
at each length is repeated in order to ensure randomization of the
unitaries and their associated implementation errors.  For our
two-qubit benchmarking demonstration, we used the set of lengths
$\{1,2,3,4,5,6\}$ and generated between $15$ and $55$ random sequences
of each length. The variation in numbers of sequences is explained
below.  We implemented $100$ runs for each sequence to determine its
probability of error~\footnote{Due to an undetermined problem in the
  control code, for approximately $1/20$ of experiments, the record
  for one run is missing.  Thus, for experiments with nominally $100$
  runs, occasionally only $99$ runs were recorded.}.

The experimental runs yield an average probability of error $E(l)$ for
each length $l$, where the average is over the sequences of this
length and their runs.  To analyze $E(l)$ we start by making the
simple assumption that each step's error behaves as a completely
depolarizing channel (see, for example, Ref.~\cite{Nielsen2001},
pg. 378) characterized by error probability $\varepsilon_g$
independent of its gates or position in the sequence. Similarly, we
assume an overall error probability $\varepsilon_m$ for state
preparation, the last inverting gate and its Pauli randomization, and
measurement. Then the mean of $E(l)$ with respect to repetitions of
the experiment satisfies
\begin{equation}
\bar E(l) = \frac{3}{4}\left(1-(1-4\varepsilon_m/3)
                               (1-4\varepsilon_g/3)^l\right)
\label{Eq1}
\end{equation}
for two qubits. The case of more than two qubits is discussed in
Sec.~\ref{sec:benchn}.  Note that $4\varepsilon/3$ is the
depolarization probability if the error probability is $\varepsilon$.
The probability of not depolarizing the state in a sequence is the
product of the probabilities of not depolarizing in each step.  To
derive the equation, note that in terms of $\bar E(l)$, the
probability of not depolarizing is $1-\frac{4}{3} \bar E(l)$.

Assuming that the experimental observations are consistent with the
simple exponential behavior suggested by this formula, we use it as
the defining formula for the EPO $\varepsilon_g$ of a random Clifford
unitary, regardless of the actual behavior of errors.  In particular,
in the context of these benchmarks, we associate the EPO with the
decay parameter of the error probabilities $E(l)$ rather than a
particular exact parameter of the underlying physical errors. This
supports the platform-independent use of randomized benchmarking.  If
the simple depolarizing assumption does not hold, then $\bar E(l)$ may
exhibit non-exponential and transient behaviors; see the discussion
below. However, the twirling effected by the randomization is intended
to induce behavior that matches the one implied by this assumption.

To isolate the EPG of the phase gate $\hat{G}$ (or any other gate) we
generate a second set of sequences by inserting $\hat{G}$ after each
random Clifford unitary. The final inverting Clifford unitary is
chosen in the same way as before, taking into account the effect of
the additional $\hat{G}$ gates to ensure that the final state is a
predictable computational basis state in the absence of errors.  Under
the same idealizing assumptions that yield Eq.~\eqref{Eq1}, the
average probability of error $E'(l)$ measured for the implementation
of this experiment satisfies Eq.~\ref{Eq1}, but with a different value
of $\varepsilon_g$ due to the additional operation in each step.  In
an ideal experiment $\varepsilon_m$ should be the same, but the model
must take into consideration that it might have changed, for example
due to experimental drifts.  Explicitly,
\begin{equation}
\bar E'(l) =
\frac{3}{4}\left(1-(1-4\varepsilon'_m/3)
                   (1-4\varepsilon'_g/3)^l\right),
\label{Eq2}
\end{equation}
where $\varepsilon'_g$ is the probability of error of a step
consisting of a random Clifford gate and $\hat{G}$. In this
context, the assumptions on the error behavior of $\hat G$ could be
relaxed from simple depolarization.  We can isolate the EPG
$\varepsilon_{\hat{G}}$ of $\hat{G}$ by solving the identity
\begin{equation}
(1-4\varepsilon'_g/3) = (1-4\varepsilon_g/3)(1-4\varepsilon_{\hat{G}}/3),
\label{Eq3}
\end{equation}
which gives
\begin{equation}
\varepsilon_{\hat{G}} =
\frac{3}{4}\left(1-\frac{1-4\varepsilon'_g/3}{1-4\varepsilon_g/3}\right).
\label{Eq4}
\end{equation}

It is helpful to run randomized benchmarks on subsets of the available
qubits so that results can be compared to other experimental platforms
that have different numbers of available computational qubits and for
investigating differences in behavior that depend on (for example)
geometrical relationships between qubits.  If possible, these
benchmarks should be run in parallel on disjoint subsets.  For these
reasons, we checked the performance of the one-qubit gates in parallel
on the two ion qubits. Because of the pre-existing benchmarks, we did
not implement the above protocol for each qubit, but used a
one-qubit benchmarking protocol similar to that of Ref.~\cite{Knill08}. Briefly, the length of a sequence is the number of steps that consist of a Pauli
gate ($\pi$-pulse) followed by a Clifford gate ($\frac{\pi}{2}$-pulse)
on each qubit. Each step can be thought of as implementing a random
computational gate.  The gate sequence is followed by a Pauli gate and
Clifford gate chosen to yield a predictable measurement outcome in the
$Z$ basis for each qubit. The Pauli gates are chosen with equal
probability to be rotations about the $\hat{x}, \hat{y}$ or $\hat{z}$
axis or the identity.  The Clifford gates are chosen with equal
probability from the following five options: $\hat{R}_x(\pm \pi/2),$ $\hat{R}_y(\pm \pi/2),$ or the identity.  When many subsequent gates are composed together, this distribution of Clifford gates demonstrates
favorable convergence to a uniformly random Clifford unitary
 in comparison with the distribution in Ref.~\cite{Knill08}.  The introduction of identity gates into the Clifford gate step reduces the average expected number of $\frac{\pm\pi}{2}$-pulses in that step from $1$ to $0.8$.

\section{Experimental Implementation}
\label{sec:experiment}

\begin{figure}
\includegraphics[width=\linewidth]{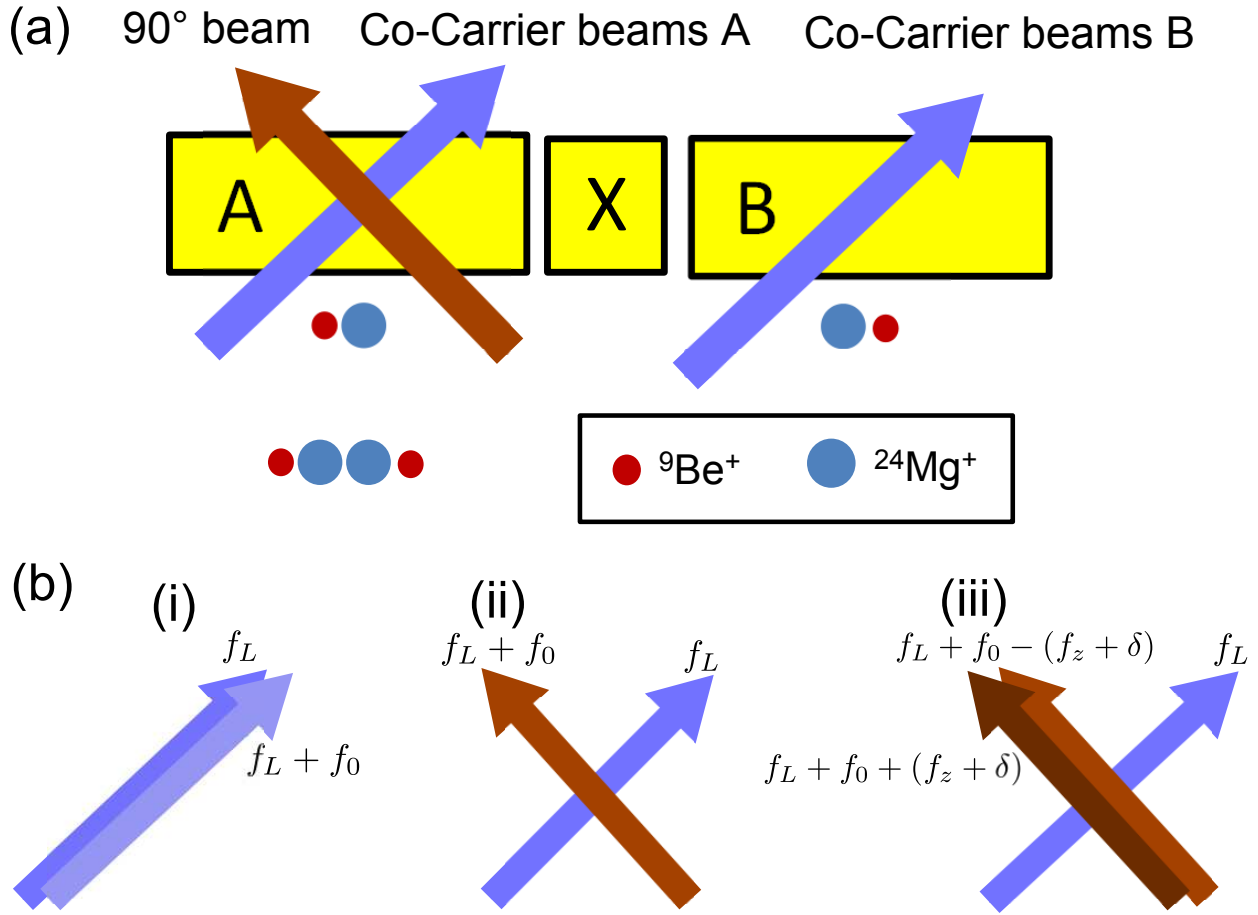}
\caption{\textbf{Experimental setup.}  (a) Schematic showing the two
  trapping zones, ion positions, and laser beam paths used (not to
  scale).  Ions are trapped in trap zones A and B.  An electrode X
  between these trap zones is used to separate and recombine
  ions~\cite{Jost09,Home09,Hanneke10}.  To perform one-qubit rotations
  the ions are separated such that a $^9$Be$^+\mathbf{-}^{24}$Mg$^+$
  pair is trapped in each zone (depicted directly under the
  electrodes). To perform the entangling gate, all four ions are
  combined in trap zone A (bottom left).  The beam waists are
  approximately $25\siunit{$\mu$m}$ in the vertical direction and
  $30\siunit{$\mu$m}$ along the trap axial direction, which is large
  compared to the extent of the two-ion and four-ion crystals
  ($6\siunit{$\mu$m}$ and $11\siunit{$\mu$m}$). (b) Laser beam
  configurations and frequencies used for different operations. (i)
  Two co-propagating beams induce Raman carrier transitions in either
  trap zone used for single-qubit gates. An acousto-optic deflector is
  used to direct the co-carrier to either trap zone. (ii) The
  90$^{\circ}$ beam is directed to trap zone A at 90$^{\circ}$ with
  respect to the co-carrier beam paths such that the wave-vector
  difference is along the trap axis. These beams induce carrier
  transitions used as part of the phase gate $\hat{G}$. (iii) Three
  beams induce the M\o lmer-S\o rensen gate used as part of
  $\hat{G}$. Beams with different frequencies depicted as slightly
  displaced arrows are actually overlapped in the experiment.  Details
  and frequency definitions are provided in the text.}
\label{trapfig}
\end{figure}

We perform the benchmarking demonstration with the ion-trap system
described in~\cite{Jost09,Home09,Hanneke10} using updated
techniques. This system includes most of the features of the scalable
quantum computing architecture of~\cite{Bible,Kielpinski02}. We trap
four ions in a six-zone linear Paul trap: two $^9$Be$^+$ ions that
serve as the qubits, and two $^{24}$Mg$^+$ ions that are used for
sympathetically recooling the qubit ions during the sequences.  The
ions form a linear chain along the axis of the trap, which is the axis
of weakest confinement. The two-qubit phase gates are performed with
all four ions in the same trap zone, in the order
$^9$Be$^+\mathbf{-}^{24}$Mg$^+\mathbf{-}^{24}$Mg$^+\mathbf{-}^9$Be$^+$
(Fig. \ref{trapfig}a bottom left). Individual addressing of the ions
for one-qubit rotations is achieved by separating the ions into two
trap zones $0.37\siunit{mm}$ apart with a single
$^9$Be$^+$-$^{24}$Mg$^+$ pair in each zone (Fig. \ref{trapfig}a below
electrodes).

The qubit states are the $|F = 1, m_F = 0
\rangle\equiv|\uparrow\rangle$ and $|2,1
\rangle\equiv|\downarrow\rangle$ hyperfine states of $^9$Be$^+$, where
$F$ and $m_F$ are the total angular momentum quantum numbers. The
energy difference between these states is first-order insensitive to
magnetic-field fluctuations at the applied field of
$0.011964\siunit{T}$~\cite{Home09,Hanneke10}. At the beginning of each
experiment we prepare the $^9$Be$^+$ ions in the $|\downarrow
\downarrow\rangle$ state.  At the end of each experiment, we detect
the qubit states by transfering the $|\downarrow \rangle$ and
$|\uparrow \rangle$ states to the $|2,2 \rangle$ and $|1,-1 \rangle$
states, respectively and then apply a $\sigma_+$-polarized laser beam
that is directed to trap zone A (Fig. \ref{trapfig}) and resonant with
the S$_{1/2}$ $|2,2 \rangle$ $\leftrightarrow$ P$_{3/2}$ $|3,3\rangle$
cycling transition. The presence (absence) of ion fluorescence
observed with a photomultiplier tube indicates the
$|\downarrow\rangle$ ($|\uparrow\rangle$) state.  For a single
$^9$Be$^+$ ion, the average number of photons collected in
$250\siunit{$\mu$s}$ is typically $30$ for the $|2,2 \rangle$ state
and $1.5$ for the $|1,-1 \rangle$ (limited by stray light).  This
allows us to analyze each detection individually with a threshold
detection level of around $11$ counts.  To measure both qubits we
first detect the state of the left qubit while the other is held in
trap zone B (Fig. \ref{trapfig}).  Then we optically pump the left
qubit to the $|2,2 \rangle$ state and transfer it into the ``dark''
$|1,-1 \rangle$ state. Finally, we bring both qubit ions into trap
zone A and apply the same procedure to detect the state of the right
qubit.

One-qubit rotations about a vector in the $x-y$ plane are implemented
with the ``co-carrier'' laser beams (Fig. \ref{trapfig}) that cause
stimulated-Raman $|\downarrow \rangle \leftrightarrow |\uparrow
\rangle$ transitions on the $^9$Be$^+$ ions after they are separated
and held in different trap
zones~\cite{Jost09,Home09,Hanneke10}. Specifically, the carrier
transitions perform $\hat{R}(\theta,\phi) = e^{-i
  \frac{\theta}{2}\sigma_{\phi}}$, where $\sigma_{\phi} =
\cos(\phi)\sigma_x+\sin(\phi)\sigma_y$, and $\phi$ depends on the
phase difference of the laser beams at the position of the ion(s).
For co-carrier transitions (on-resonance qubit transitions), two
co-propagating laser beams have frequencies $f_{L}$ and $f_{L}+f_0$
(see Fig.~\ref{trapfig}b (i)\ ), where $f_{L}$ is the principal laser
frequency at $957.132$ THz, which is approximately $70$ GHz below the
$S_{1/2}$ to $P_{1/2}$ transition frequency and $f_0 = 1.207353$ GHz
is the qubit transition frequency.  The co-carrier laser beam position
along the trap axial direction is controlled with an acousto-optic
deflector that allows the beam to address ions in either trap
zone. The pulse duration for a single-qubit rotation by $\pi$ and
$\frac{\pi}{2}$, using the co-carrier beam configuration shown in
Fig. \ref{trapfig}b (i), was approximately $9$ and $4.5$ $\mu$s,
respectively.  One-qubit $\sigma_z$ gates ($\hat{R}_z(\phi) = e^{-i
  \frac{\phi}{2}\sigma_z}$) are implemented in software by shifting
the RF phase of all future rotations for that qubit by $-\phi$.
Identity gates are implemented with a wait time equal to
$4\siunit{$\mu$s}$.

One laser beam propagating along the co-carrier beam path and a pair
of laser beams propagating along the 90${^\circ}$ beam path
(Fig. \ref{trapfig}) are used to implement phase gates $\hat{G}$. In
contrast to the experiments in \cite{Jost09, Home09, Hanneke10}, which
implemented two-qubit $\hat{G}$ phase gates directly on hyperfine
states, we use one-qubit rotations and a M\o lmer-S\o rensen (MS)
gate~\cite{Sorenson99} to implement $\hat{G}$~\cite{Lee05}.  In the
previous experiments, implementation of the phase gate required that
the qubit ions' states be transferred from the qubit manifold, where
the qubit frequency is first-order independent of magnetic-field
fluctuations, to other hyperfine
states~\cite{Home09,Hanneke10}. However, the MS gate can be performed
directly on the qubit states.  To implement $\hat{G}$ we surround a MS
gate pulse, $\hat{U}_{MS} = e^{\frac{-i
    \pi}{4}\sigma_{\phi}^{(1)}\sigma_{\phi}^{(2)}}$, with carrier
$\pi/2$-pulses on both ions by use of two laser beams as shown in
Fig. \ref{trapfig}b (ii) and (iii).  The resulting three pulse
sequence is
$e^{-i\frac{\pi}{4}(\sigma_{\phi_+}^{(1)}+\sigma_{\phi_+}^{(2)})}\hat{U}_{MS}e^{-i\frac{\pi}{4}(\sigma_{\phi_-}^{(1)}+\sigma_{\phi_-}^{(2)})}
= \hat{G}e^{-i\frac{\pi}{4}}$, where we use $\phi_{\pm} = \phi \pm
\frac{\pi}{2}$ and where the overall phase factor after $\hat{G}$ has
no physical consequence in this setting.  The advantage of using
$\hat{G}$ as our elementary two-qubit gate rather than the MS gate is
that this three-pulse sequence is insensitive to slow changes in the
optical path-length difference of the non-copropagating beams, which
cause $\phi$ to change~\cite{Lee05}.  The duration of the
$\hat{U}_{MS}$ pulse is $20\siunit{$\mu$s}$ and the duration of each
carrier $\pi/2$ pulse using the beam configuration shown in
Fig. \ref{trapfig}b (ii), is approximately $1.5\siunit{$\mu$s}$. Due
to wait periods between pulses that are necessary to stabilize the
feedback loops that control the laser pulse amplitudes and phases, the
three pulse sequence requires $110\siunit{$\mu$s}$ to complete.
Before performing each $\hat{G}$ gate we sympathetically laser-cool
the four-ion crystal, first using Doppler and then Raman sideband
cooling of the $^{24}$Mg$^+$ ions~\cite{Jost09, Home09, Hanneke10}.
This ensures that each time we implement $\hat{G}$, the motional modes
along the axial direction are cooled to near the ground state. The
cooling light interacts only with $^{24}$Mg$^+$ and thus preserves the
qubit state coherences.

\begin{figure}
\includegraphics[width=\linewidth]{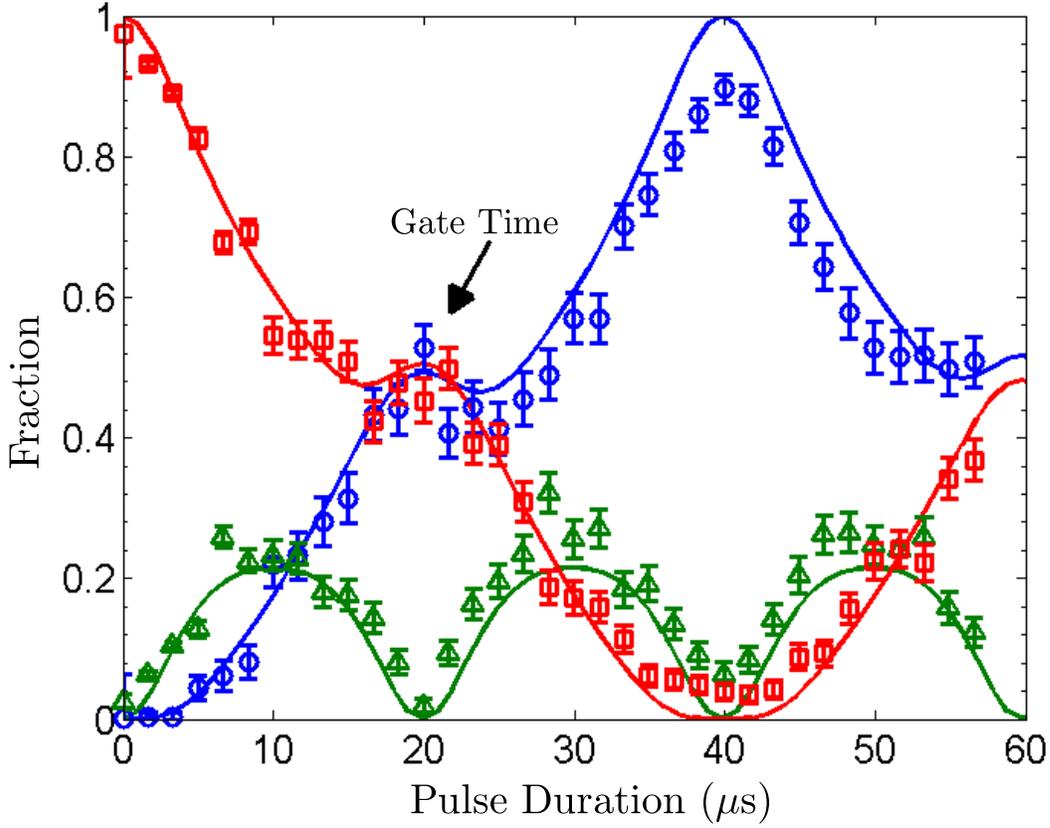}
\caption{\textbf{M\o lmer-S\o rensen gate.}  We simultaneously apply
  detuned red and blue sideband MS-gate beams to an initial
  $|\downarrow\rangle|\downarrow\rangle$ state for varying durations
  and observe the frequency with which we find both ions in the
  $|\downarrow\rangle$ state (red squares), both ions in the
  $|\uparrow\rangle$ state (blue circles), or one ion in each state
  (green triangles).  From these curves, we can determine the gate
  time ($\frac{1}{\Delta f}$) for the MS gate. Here it is
  approximately $20$ $\mu$s, at which point the qubit states are
  entangled and ideally in the state
  $\frac{1}{\sqrt{2}}(|\downarrow\rangle|\downarrow\rangle+e^{i\phi}
  |\uparrow\rangle|\uparrow\rangle)$, where $\phi$ depends on the
  phases of laser beams at the ions' position and can vary from
  experiment to experiment (see text).  The solid lines show the
  theoretical results for an ideal gate.  To perform a phase gate we
  surround the MS-gate pulse with two $\frac{\pi}{2}$-pulses by use of
  the laser beams as depicted in Fig.~\ref{trapfig}b. (ii).  The
  points and their error bars were determined by photon-count
  histogram fitting from $250$ runs.}
\label{MSGate}
\end{figure}

In more detail, the MS gate requires the simultaneous application of
detuned blue and red sidebands.  To achieve this, we overlap three
laser beams with different frequencies in trap zone A
(Fig.~\ref{trapfig}b (iii)\ ).  One laser beam propagates along the
co-carrier beam path with frequency $f_{L}$. The other two beams
co-propagate in the 90$^{\circ}$ beam path at frequencies of
$f_{L}+f_0 \pm (f_z+\delta)$, where $f_z$ is the frequency of a
motional mode and $\delta \ll f_z$ is a detuning. The two laser beams
are derived from a single beam that is split and passed through
different double-pass acousto-optic deflectors such that they end up
with a frequency difference of $2(f_z+\delta)$.  The split beams are
then recombined on a $50$-$50$ beam splitter with one port directed to
the ions and the other going to a photo-detector that is used to
measure and stabilize the phase of the beat note, as required to
realize $\hat{U}_{MS}$ \cite{Lee05}.  To implement $\hat{G}$, we
simultaneously address the two highest-frequency axial motional modes
for the four-ion crystal at $f_z = 5.487\siunit{MHz}$ and $f_z' =
5.739\siunit{MHz}$~\cite{Jost09}.  The detuning $\delta$ must be
chosen such that the detuning from one mode is an integer multiple of
the detuning from the other in order to fully disentangle both
motional states from the qubit states at the end of the gate.
Experimentally, we found a detuning of $\delta = 50\siunit{kHz}$ above
$f_z$ was optimal given our laser beam intensities, which implies that
the MS gate was implemented with one phase-space loop on the $f_z$
mode and four loops on the $f_z'$ mode~\cite{Sorenson99}.  In
Fig. \ref{MSGate} we plot the observed fraction of both ions in the
$|\downarrow\rangle$ state (red squares), both ions in the
$|\uparrow\rangle$ state (blue circles), and one ion in each state
(green triangles) as a function of the duration of the red and blue
sideband pulses applied to an initial state of
$|\downarrow\rangle_1|\downarrow\rangle_2$.  The MS gate is completed
in approximately $20\siunit{$\mu$s}$ ($\frac{1}{\delta}$).

For the Clifford and phase-gate benchmarks, we generated random
sequences of lengths $1,2,3,4,5,6$.  The respective number of
sequences implemented was $45, 55, 53, 39, 28, 15$ for the Clifford
benchmark, and $46, 54, 53, 38, 28, 15$ for the phase-gate benchmark,
in order of sequence length.  Each time we performed a sequence for
the Clifford benchmark we then immediately performed the corresponding
phase-gate benchmark sequence. Each sequence
was implemented $100$ times. From the measurement outcomes, we
determined the fraction of measurements that matched the prediction.
The data shown in Fig.~\ref{RBdata} was obtained in four successive
sets of experiments on the same day.  During and between the sets we
periodically recalibrated the magnetic field and the laser frequencies
needed for sympathetic cooling of $^{24}$Mg but not the other pulse
parameters.  Within each set, the sequences were randomized with
respect to length.  However, the first (second, third, fourth) set
involved sequences of lengths $1$ to $3$ (to $4$, $5$, $6$,
respectively). In particular, sequences of length $6$ were run only in
the last set of experiments, which is why there are fewer sequences of
length $6$ contributing to the data. During the experimental runs for
these sets, we observed one ion-loss event. We did not implement
longer sequences because ion-loss events became a problem for lengths
greater than $6$.

In our implementation, a Clifford unitary took $4.5\siunit{ms}$ on
average. For the sequences with an extra $\hat{G}$ gate inserted after
each step, $7.5\siunit{ms}$ per step was typical. The most
time-consuming elements of the sequence implementations were the
sympathetic recooling of the ions after each recombination of the ions
into trap zone A, followed by the separation and recombination
processes.  Each sequence began with approximately $10\siunit{ms}$ for
state preparation and laser lock stabilization.  Thus, a sequence of
length $6$ with an extra $\hat{G}$ inserted at each step lasted
approximately $55\siunit{ms}$.  Longer sequences resulted in an
accelerated rate of ion loss events (on the order of a loss event per
minute), which can likely be attributed to a decreased probability of
recovery from background gas collisions that can occur at any point
during the sequences.  Before running each sequence, two warmup
sequences with $100$ experiments each were run to make sure the
experiment was in a steady state; the results of these experiments
were not recorded.  Switching from one sequence to the next required
$3\siunit{s}$ to $4\siunit{s}$ of computer time to reprogram the
control hardware.  In total, all of the Clifford benchmarks, including
the sequences used to benchmark $\hat{G}$, were completed in
approximately $1$ hour and $45$ minutes, which also includes the time
durations needed for periodic recalibrations of the magnetic field as
well as the time period to reload a set of ions following the only
ion-loss event.

The parallel one-qubit benchmark whose results are shown in
Fig.~\ref{RBdataSQ} was executed in one set, after all of the
two-qubit benchmarks and following a recalibration of the one-qubit
gates.  The number of sequences implemented was $15,13,6,13,12,14$ for
sequence lengths of $2,3,4,6,8,12$, respectively.  We ran each sequence
$100$ times, as before. In order to approximately replicate the
conditions of the experiment for the two-qubit benchmark, in each
step, the ions were recombined into a single trap zone, recooled and
then held for approximately the same duration required to execute
$\hat{G}$ before being separated again for the next sequence step.

\section{Experimental Results}
\label{sec:results}

\begin{figure}
\includegraphics[width=\linewidth]{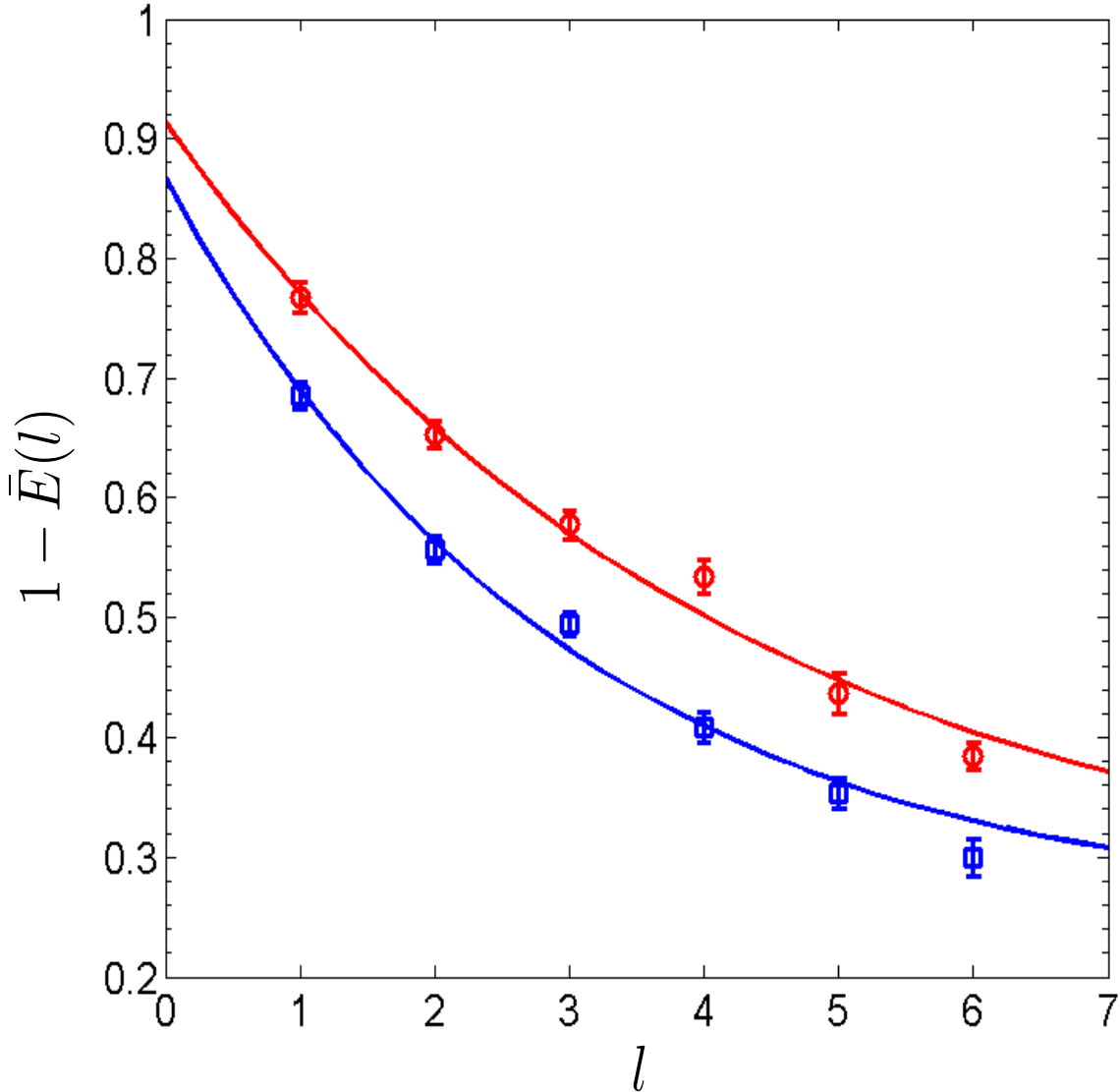}
\caption{\textbf{Randomized benchmarking of two-qubit gates.}  The red
  circles show one minus the average probability of measuring an error
  at the end of sequences of random Clifford unitaries $\bar{E}(l)$ as
  a function of the sequence length $l$.  By fitting the data to the
  expression in Eq.~\eqref{Eq1} (red line), we find an error per
  random Clifford unitary $\varepsilon_g = \epomean\epoerrp$. The
  preparation/measurement error, $\varepsilon_m$, is
  $\epoprepmean\epopreperrp$ (recall that measurement error includes
  the error for an additional inverting gate before detection).  Blue
  squares show the results for running random sequences with an
  additional $\hat{G}$ inserted after each step.  Fitting this data to
  Eq.~\eqref{Eq2} yields an error of $\varepsilon_{\hat{G}} =
  \epgmean\epgerrp$. In this case the preparation/measurement error,
  $\varepsilon_m'$, is $\epgprepmean\epgpreperrp$.  The error bars in
  the plot represent the standard deviation of the mean of the
  sequences' frequency of correct measurement outcome.  Error bars for
  inferred parameters are based on bootstrap resampling; see the
  text. }
\label{RBdata}
\end{figure}

The red data points and curve in Fig.~\ref{RBdata} show the results
from the experimental Clifford gate benchmark and their match to an
exponential decay. The match gives a Clifford unitary EPO of
$\varepsilon_g = \epomean\epoerrp$.

The blue data points and curve in Fig.~\ref{RBdata} show the results
from the $\hat{G}$ benchmark. Curve-fitting and solving the above
equations for $\varepsilon_{\hat{G}}$ give an EPG of
$\varepsilon_{\hat{G}} = \epgmean\epgerrp$.

\begin{figure}
\includegraphics[width=\linewidth]{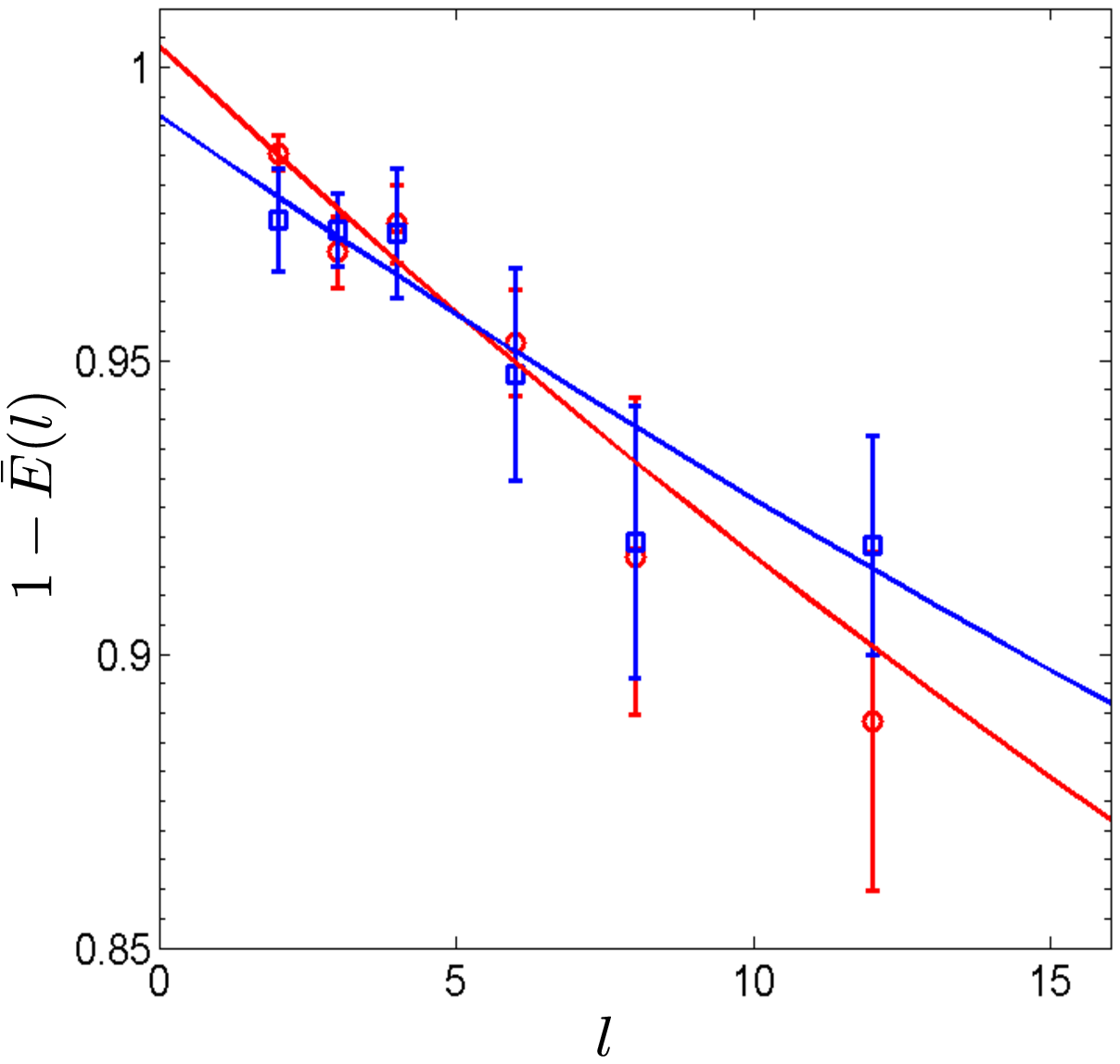}
\caption{\textbf{Randomized benchmarking of one-qubit gates.}  Red
  circles and blue squares show one minus the average probability of
  error for each qubit independently. The solid lines are the best
  fits of the data to $\bar{E}(l) = \frac{1}{2}-
  \frac{1}{2}(1-2\varepsilon_m)(1-2\varepsilon_g)^l$ where $l$ is the
  sequence length, $\varepsilon_m$ is related to the state preparation
  and readout fidelities of the two qubits, and $\varepsilon_g$ is the
  error per step in the sequence.  We find the errors
  per step to be $\epgoneleftmean\epgonelefterrp$ and
  $\epgonerightmean\epgonerighterrp$, respectively. }
\label{RBdataSQ}
\end{figure}

We determined the errors per step on each qubit independently with the
parallel one-qubit benchmarks explained above. The results from the
benchmarks are shown in Fig.~\ref{RBdataSQ}.  The inferred one-qubit
errors per step are $\epgoneleftmean\epgonelefterrp$ and
$\epgonerightmean\epgonerighterrp$ for the respective qubits.
Using the assumption that laser pulses dominate the error per step, 
these results can be compared to the protocol of Ref.~\cite{Knill08}
through multiplication by the ratio $2:1.8$ of $\frac{\pm\pi}{2}$-pulses per step in the two protocols.

\section{Data Analysis Methods}
\label{sec:analysis}

In the limit of very large numbers of sequences for each length, we
can use a simple, nonlinear, weighted-least-squares fit of
Eq.~\eqref{Eq1} to the fidelity curves as a function of length. The
weights are determined by the standard error of the mean for the
fidelity at each length.  Note that some non-linear least-squares fitting
functions compute the error in the inferred parameters from the
fitting error and ignore the scale of the errors implied by the
weights given to the individual points. Because we already have a good
estimate of the standard errors of these means, a better estimate of
the error in the inferred parameters can be obtained by direct
propagation of errors, particularly when there are few points.

For smaller numbers of sequences for each length, we must
consider that we know little about the distribution of the fidelities
for different random sequences of a given length. This distribution is
affected not only by the differences in actual pulses applied, but
also by factors such as the amount of coherence in error (see
below). However, the estimate of a given sequence's fidelity from the
$100$ experimental runs is binomially distributed. Thus we used a
partially parametric bootstrap~\cite{Efron1993} procedure to determine
a standard error for the parameters inferred by fitting.  Let $n_l$ be
the number of different sequences of length $l$ used in the
experiment.  Denote the experimentally measured fidelity of the $j$'th
sequence of length $l$ as $F(l,j)$, which is the fraction of times the
correct result was obtained during the $100$ runs of the $j$'th
sequence.  We generated artificial data for each bootstrap resample as
follows.  For each length $l$, we constructed
$F_r(l,k),k=1,\ldots,n_l$ by letting $F_r(l,k)$ be a random element of
the sequence of fidelities $F(l,j),j=1,\ldots,n_l$, picked
independently (with replacement, that is, the same element can be
picked multiple times) for different $k$.  For each $F_r(l,k)$, we
generated a random $F'(l,k)$ according to a binomial distribution, so
that $F'(l,k)$ was the fraction of $1$'s in $100$ random instances of
a $0/1$ variable where the probability of $1$ was $F_r(l,k)$. We then
averaged the $F'(l,k)$ for each length $l$ and computed the fit in the
same way as it was computed for the real data to obtain the inferred
parameters for this resample.  This resampling procedure was repeated
$1000$ times, yielding $1000$ resampled values for each of the
parameters.  The standard errors in the reported parameters are
determined as the square-root of the variances of the corresponding
resampled parameters. The values of the parameters are still the ones
from the fit of the original experimental data, as this is the
less biased estimate and requires no bootstrapping.

\section{Consistency Checks}
\label{sec:consistency}

Although the relationship
between EPOs and physical errors in gates is not known in general,
specific benchmarking protocols provide well-defined EPOs that can be
compared across platforms.  However, for an implementation of the
benchmark to be convincing, there are several assumed or expected
properties that can be checked.  These include the following: We can
determine whether or not the fidelity curves are consistent with a
simple exponential as a function of sequence length, and if not,
analyze the deviations.  Given that we know the implementations of the
Clifford unitaries, we can compare the EPO for a Clifford unitary with
that inferred from the EPG for a phase gate and the one-qubit
benchmark.

First we consider the exponential fits shown in Fig.~\ref{RBdata}.
The $\chi^2$ values (four degrees of freedom) for the two curves are
$9.28$ and $9.48$, respectively.  The higher one corresponds to a
$p$-value of $0.0501$, approximately the conventional boundary for
significance.  There is other evidence that the exponential model may
not be a good fit.  First, the two preparation/measurement errors are
expected to be the same, but the fits seem to suggest otherwise,
although the statistical significance is not strong. Second, both sets
of data seem to dip below the fit near the end. Together these
observations suggest an increased EPO for later Clifford
unitaries. Indeed, dropping the first points from the analysis
suggests higher EPOs. For example, the fits for the last four and
three sequence lengths have EPOs of
$\varepsilon_g=\epoCtoF\epoCtoFerrp$ and
$\varepsilon_g=\epoDtoF\epoDtoFerrp$, respectively. The corresponding
EPGs are $\epgCtoF\epgCtoFerrp$ and $\epgDtoF\epgDtoFerrp$.  Note that
the second values are from a two-parameter fit to three points, which
reduces their significance.  The fits to the first four sequence
lengths give an EPO of $\epoAtoD\epoAtoDerrp$ and an EPG of
$\epgAtoD\epgAtoDerrp$. The higher EPOs for longer sequences could be
related to the fact that unlike the shorter ones, they were run only
in the last or last two sets of experiments, without recalibrating
pulse parameters except as noted above. We attempted to confirm this
hypothesis by analyzing the results for sequence lengths one to three
separately for each set. The results of this analysis are consistent
with a drop but have insufficient signal-to-noise to be conclusive.

In principle, we would like the platform to have the property that
errors reach stationary behavior soon after state preparation, and the
benchmark's reported EPO should reflect the stationary error. As noted
above, we were not able to implement long enough or sufficiently many
sequences to clearly observe stationary behavior, or to determine the
extent to which the behavior is nonstationary.  The EPO and EPG
reported in the abstract are determined from all six lengths
tested---given the ``early'' and ``late'' values above, we believe
that they are a good representation of mid-length behavior of gate
errors.  Our inability to consistently run sequences of length longer
than six prevents any claims of stationary behavior.

Other issues with the exponential fits that can arise include the
possibility that the curves are a mixture of exponentials, as would be
expected if the EPOs change slowly compared to the time required to
run a sequence.  In this case, the apparent EPOs would tend to
decrease with increasing sequence length, as the higher-EPO runs affect
the loss at shorter lengths, but tail behavior is dominated by the
slowest decay.  Given the observations of the previous paragraph and
the available statistics, we cannot usefully test for this
possibility.

Now we consider consistency between the measured EPOs and EPGs.  We
estimate the EPO that we should have measured given the EPGs obtained
from the one-qubit and the phase gate benchmarks.  For this estimate,
we count the complexity of sequences of one-qubit pulses in terms of
the number of effective $\frac{\pi}{2}$-pulses applied. This counts
only pulses around the $\pm \hat x$ or $\pm \hat y$ axes, taking into
consideration that $\hat z$-axis pulses and identity gates are
essentially error-free. The $\pi$-pulses are counted as two
$\frac{\pi}{2}$-pulses. Coherent error addition is neglected.  The
one-qubit benchmark's steps each have an average of $1.8$ effective
$\frac{\pi}{2}$-pulses per qubit. If we use $0.0085$ as a
representative error probability per step from the one-qubit benchmark
(Fig.~\ref{RBdataSQ}), we obtain $e_1(1)=(6/5)*0.0085/1.8=0.0057$ as the
linearized error probability per one-qubit $\frac{\pi}{2}$-pulse.  The
factor of $6/5$ converts the average probability of error for one
qubit to that for two qubits under the assumptions that the other
qubit has no error and any Pauli error is twirled to a depolarizing
error.  For the purpose of this calculation, we take the error
probability per phase gate to be $0.069$. Each step in the Clifford
benchmark has an average of $6.5$ effective $\frac{\pi}{2}$ pulses and
$1.5$ phase gates.  The linearized error probability for a step can
therefore be estimated as $1.5*0.069+6.5*0.0057=0.14$, with a
standard error of about $0.02$, if we add the statistical errors in
quadrature. This linear approximation is expected to give a
pessimistic estimate, but in this case, the nonlinear correction is
smaller than the error in the estimate. While our estimate gives a
value below the measured EPO, the difference appears not to be
statistically significant. We emphasize that the above strategy for
estimating the EPO from EPGs neglects coherent error addition, which
tends to increase the error, and internal error cancellation that
could arise from the way pulses are combined within a step.

\section{Estimates of Physical Sources of Error}
\label{sec:estimates}

We consider known sources of errors and estimate their contribution to
the EPG of $\hat{G}$. Spontaneous emission is a fundamental source of
error for transitions driven by stimulated-Raman transitions; here the
laser beams are tuned approximately $70\siunit{GHz}$ below the
$P_{1/2}$ state~\cite{Ozeri07,Uys2010}.  We simulate that for our
laser parameters, this should contribute an error probability of
$0.001$ to a one-qubit $\pi$-pulse and $0.013$ to the phase
gate. (Recall that the phase gate consists of an MS gate surrounded by
$\pi/2$-pulses.)

\begin{figure}
\includegraphics[width=\linewidth]{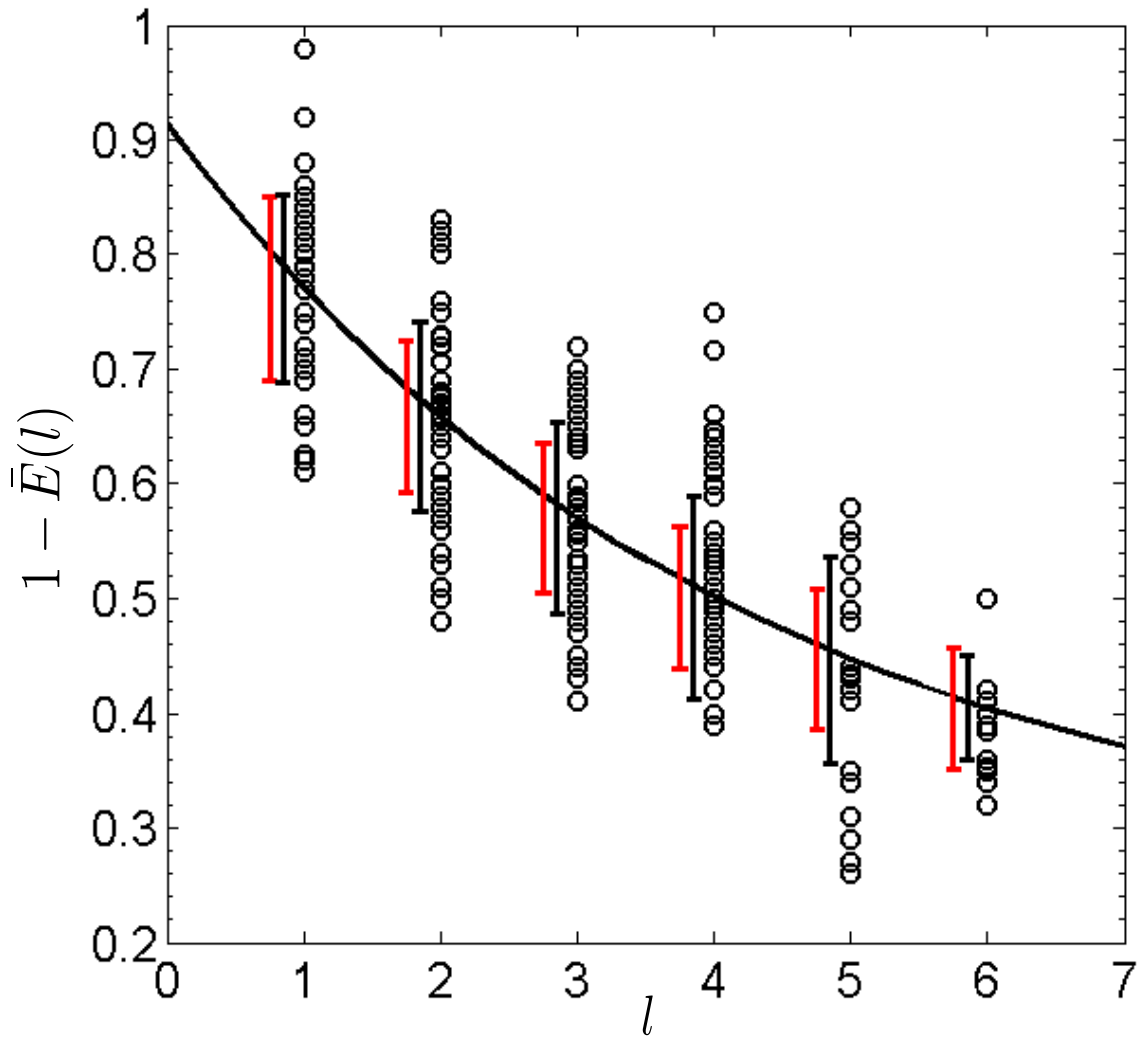}
\caption{\textbf{Scatter in the error for randomized benchmarking
    data.} The open circles show one minus the experimentally measured
  average probabilities of error for the individual sequences of
  random Clifford unitaries as a function of sequence length for the data
  shown in red in Fig.~\ref{RBdata}.  The total numbers of sequences
  shown at sequence lengths $1,2,3,4,5,6$ are $44,54,52,38,28,15$,
  respectively.  The average error is the percentage of times at least
  one of the qubits was not found in the expected state in $100$
  experiments.  The red error bars to the left of the data at each
  length show the expected standard deviation if the error variation
  is due to variation in the number of phase gates needed to implement
  the random Clifford unitaries used in the sequences and the binomial
  statistics for the $100$ runs for each sequence. The black error
  bars show the standard deviation of the set of fidelities measured
  for the corresponding length. The solid line is the fit to
  Eq.~\ref{Eq1}. }
\label{alldata1}
\end{figure}

Errors can also arise from imperfect calibrations and slow drifts of
the gate parameters.  These parameters include beam intensities,
frequencies, phases, and pulse durations. These types of errors are
coherent in the sense that for any given run, each implemented gate
still causes a unitary change in state, but not exactly the intended
one.  To determine whether such errors contribute significantly to the
measured EPOs, we consider the variation in fidelities for different
sequences of a given length. Coherent error contributions typically
result in a variation that is larger than that expected from a simple
statistical analysis~\cite{Knill08}.  For our experiment and in the
absence of coherent errors, we attribute the largest sources of variation
in the fidelities of Fig.~\ref{RBdata} to the varying number
of $\hat{G}$ gates and single-qubit rotations needed to implement each
step's random Clifford unitary, and to the binomial statistics for
the fidelities inferred from the $100$ runs of each sequence.
Fig.~\ref{alldata1} compares the actual variation and the
variation predicted from the statistics of the number of phase gates
per step and the binomial statistics. We did not include the variation
in numbers of one-qubit gates due to their significantly smaller
error. The gate statistics and binomial statistics are independent, so
their contributions were added in quadrature.  The predicted variation
is generally somewhat less than the measured one but does not indicate
coherence of the errors in a given sequence because our simple model
does not account for all incoherent effects.

The main sources of coherent errors are due to drifts in beam
intensity, relative laser field phases for the beams implementing the
phase gate, and Stark-shifted frequencies.  These drifts result in
errors in pulse time, and phase and frequency calibration, each of which are
estimated to contribute approximately equally to the EPG.  We estimate
that their total contribution to the phase gate EPG is less than
$0.03$.

There are a number of lesser sources of error to consider.  In
addition to the slow intensity drifts included above, there are also
fluctuations in intensities that can be slow compared to sequence
duration but are too fast to be calibrated out.  For example, these
can arise from fluctuations in laser power or from noise in the
position of the laser beams with respect to the ions, due to vibration
and air movement.  Such fluctuations in intensity lead to loss of
visibility in Rabi flopping curves~\cite{Bible,Knill08}. From such
curves, we determined that the Rabi rate on the carrier transition
with the non-copropagating beams fluctuates by $\frac{\delta
  \Omega}{\Omega} = 0.029(1)$ from experiment to experiment.
This results in a contribution of $2 \times
10^{-3}$~\cite{Benhelm2008} to the phase gate EPG.

Due to the finite Lambe-Dicke parameter for the ions, fluctuations in
ion motional energy can cause errors in the MS gate~\cite{Bible, Sorensen00}.
As in Ref.~\cite{Jost09}, we estimate that each motional mode is
cooled to an average excitation of at most $0.2$ quanta before the
implementation of each phase gate.  This leads to errors in the MS
gate of $6\times10^{-4}$ due to the finite excitation of the modes
directly involved in the gate and $1\times10^{-3}$ due to the
fluctuating Debye-Waller factors of the other modes
combined~\cite{Bible,Sorensen00}.

Intrinsic background heating for the ions results in motional
decoherence in the MS gate while the spins and motion of the ions are
entangled.  We measure a heating rate for a single $^9$Be$^+$ ion to
be $0.3$ to $0.5$ quanta per millisecond in the common axial mode at a
confinement frequency of $2.7\siunit{MHz}$.  However, the motional
modes used for the MS gate have only a small component of the
center-of-mass motion.  Conditions here are essentially the same as
those of a previous experiment~\cite{Home09}, and imply a contribution
of less than $10^{-3}$ to the phase gate's error.

To perform the phase gate, the optical path lengths of the two
non-copropagating beams should ideally remain constant throughout the
gate's three-pulse sequence, whose duration is $110\siunit{$\mu$s}$.
Measurements of the optical beat-note between the non-copropagating
Raman beams give a linewidth of order $10\siunit{Hz}$; from this we
estimate that optical-path-length fluctuations result in a phase-gate
error of the order of $10^{-3}$, assuming the relative phase
fluctuations at the beat-note detector are the same as those
experienced by the ions.

The qubit coherence time and thus the benchmark error probabilities
are affected by fluctuations in the magnetic field and its gradients,
which cause differential frequency shifts in the qubits.
Experimentally we determine a qubit coherence time by measuring the
decay of the contrast in a two-pulse Ramsey experiment as a function
of the duration between pulses.  We find a coherence time of
$4.1\pm1.7\siunit{s}$. Due to the resolution limit of our
frequency synthesizer, which serves as a clock to keep track of the
qubit phases during the randomized benchmarking experiments, we are
systematically detuned from the qubit frequency by
$270\siunit{mHz}$. The error due to this frequency offset should be
negligible, given the typical time required for a Clifford unitary of
a few milliseconds.  We measure a magnetic field difference between
trap zones A and B of $1.5\times10^{-7}\siunit{T}$, which leads to a
systematic frequency difference between the qubits of a few
millihertz depending on the exact value of the magnetic-field.

As an independent check on phase-gate fidelity, we measured the state
fidelity for a Bell state created by use of the phase gate $\hat{G}$, as
was done in Refs.~\cite{Sackett2000,Leibfried03,Benhelm2008}.  Such
measurements were performed before and after the randomized
benchmarking data was taken. Before the benchmark we determined a Bell
state fidelity of $0.91(2)$. After the benchmark we obtained
$0.90(2)$.  These fidelities include errors due to imperfect state
initialization, detection, and three carrier $\frac{\pi}{2}$-pulses
using the co-propagating beams that are needed to prepare and analyze
the state.  A measurement of the state fidelity where the Bell state
was prepared using only the MS gate and analyzed with a single
non-copropagating carrier $\frac{\pi}{2}$-pulse, thereby removing one
non-copropagating carrier pulse and three co-propagating carrier
pulses from the measurement, gave $0.94(1)$.  The fidelities are
consistent with the EPG determined by the benchmark.

Errors for one-qubit gates implemented with copropagating beams are
likely dominated by changes in the Rabi rate~\cite{Knill08}.  An
indication of whether or not long-term drifts may have affected the
two-qubit benchmark can be obtained by comparing two one-qubit
benchmarks. The first was run immediately after the two-qubit
benchmark, without recalibrating.
The second followed recalibration of the relevant pulses and is the
one that we reported above. The first one found EPGs
$\epgoneleftmeantwo\epgonelefterrptwo$ and
$\epgonerightmeantwo\epgonerighterrptwo$ for the two qubits,
respectively, suggesting that at least the second qubit's gates may
have been in need of recalibration by the end of the two-qubit
benchmarks.

Error can also be caused by loss of ions due to background gas
collisions.  We checked for loss of ions after each sequence, and if an
ion-loss event was detected, we removed the previous sequence from the
data set.  We observed a significant increase in the rate of ion loss
events for sequences involving more than $16$ ion
separation/recombination processes---one such process is needed for
each phase gate performed.  This limited the maximum length of the
sequences used in the randomized benchmarking.  The reason for the
increase in the rate of ion loss is not understood.

In summary, the errors discussed in the previous paragraphs amount to
a phase gate EPG of about $0.048$ (linearized, incoherent error
addition) to be compared to the benchmark-determined EPG of
$\epgmean\pm\epgerr$. The EPG estimate of $0.048$ includes $0.013$ for
spontaneous emission, $0.03$ for calibration imperfections, $0.002$
for intensity fluctuations, $0.0016$ for ion motion, $0.0005$ for
motional heating, and $0.001$ for optical path length
fluctuations. The fact that our measured error is greater than our
estimated error based on known physical sources suggests that our
model of errors for the phase gate is incomplete.

\section{Multi-qubit Randomized Benchmarking}
\label{sec:benchn}

Clifford benchmarks as defined above can serve as a
platform-independent strategy for comparing the quality of quantum
operations in a computational context. For $n$ qubits, the expressions
for the EPOs and EPGs of Eq.~\eqref{Eq1} and~\eqref{Eq4} are
generalized as follows:
\begin{eqnarray}
\bar E(l) &=& \frac{2^n-1}{2^n}\left(1-(1-2^n\varepsilon_m/(2^n-1))
                               (1-2^n\varepsilon_g/(2^n-1))^l\right),
\label{Eq1n}
\\
\varepsilon_{G} &=&
\frac{2^n-1}{2^n}\left(1-\frac{1-2^n\varepsilon'_g/(2^n-1)}{1-2^n\varepsilon_g/(2^n-1)}\right),
\label{Eq4n}
\end{eqnarray}
where $G$ is a gate being characterized by insertion after each step.
These equations can be established under idealizing assumptions in the
same way as Eqs.~\ref{Eq1} and~\ref{Eq4}, after observing that for $n$
qubits, the probability that the sequence does not depolarize is
$1-\frac{2^n}{2^n-1} E(l)$.  The comparison on the basis of EPOs and
EPGs obtained from the length-dependent loss of fidelity makes sense
in the absence of significant deviations from the simple exponential-decay model. The thus-measured EPGs can be meaningfully interpreted as
true average gate errors in the idealized case where the errors are
independent, depolarizing and stationary. In general, the connection
to the actual error-behavior of elementary gates is not well
understood~\cite{Magesan11}.  Nevertheless, we believe that
sufficiently small randomized-benchmark-determined EPGs are a
good indication of gate quality that can be compared to a
general-purpose fault-tolerance threshold goal such as the
often-mentioned $10^{-4}$.  In the above, we have considered how to
take into account the indications for non-exponential decay in our
data. Generally, if there is clear evidence of non-exponential decay,
the behavior and range of observed EPOs and EPGs and the extent to
which stationary behavior was achieved need to be discussed. Given
sufficiently many sequence lengths, these ranges can be determined by
considering different-length intervals. Initial transients in error
behavior may be expected even in the case where stationary behavior is
achieved for longer sequences and can be analyzed separately.

The translation of a given Clifford unitary into a circuit of
elementary gates suitable for a given platform is up to the
experimenter, so some improvements are possible by greater efficiency
of the translation rather than higher quality gates. We consider such
``software'' improvements to be potentially as useful as strictly
``hardware''-based ones. Furthermore they are usually easier for
others to implement. However, we believe that for small
numbers of qubits, such circuit translations are already sufficiently
close to optimal for software improvements of this sort to be
self-limiting.  The individual gate benchmarks implemented by
inserting specific gates after the Clifford unitaries can show
directly how much the gate quality has improved, independent of how
the Clifford unitaries have been translated into circuits of
elementary gates.

When benchmarking $n$ qubits, we suggest that the benchmarks are
applied to different subsets of the qubits so that comparable EPOs are
obtained for $n=1,2,3,\ldots$ qubits.  We recommend that such
benchmarks be applied in parallel to disjoint subsets, if possible.
This solves the problem of comparing new results to earlier ones
involving platforms with fewer qubits.  Nevertheless, it would be
helpful to have a way of comparing EPOs for the Clifford benchmark
that is independent of the number of qubits.  One possibility is to
divide the EPO by $C(n)$, the average over Clifford unitaries of the
minimum number of controlled-not gates needed to implement them with a
circuit consisting of controlled-not gates and arbitrary one-qubit
Clifford gates. For two qubits, this normalization factor is
determined by $C(2)=1.5$, so the normalized Clifford EPO for our
benchmark is $\normepo\normepoerrp$. For three qubits we determined
$C(3)=3.51$ rounded at the last digit.  Since the number of Clifford
unitaries grows very rapidly with $n$, it may be difficult to
determine the normalization factor exactly for $n>4$. It is known that
$C(n)$ scales as
$n^2/\log(n)$~\cite{Shende2003,Aaronson2004,Patel2008}.  While this
complexity may seem relatively large at first, any viable platform
must be able to implement circuits of this size, if not much larger.
In particular, any successful demonstration of the Clifford benchmark
also establishes the ability to implement non-trivial circuits for
algorithmic purposes.

If the primary purpose of the experiment is to benchmark individual
gates by inserting them after randomized unitaries, it is desirable to
find ways to achieve sufficient randomization that are more efficient
than random Clifford unitaries. In particular, to exhibit an EPG of a
gate in this way, it suffices for the random unitaries to approximate
a so-called unitary $2$-design as explained in~\cite{Dankert06}.  Such
approximations are possible with circuits involving a logarithmic
multiple of $n$ two-qubit gates~\cite{Dankert06}. How to best
translate these theoretical ideas into a practical benchmark remains
to be determined.

For the Clifford benchmark, the strategy for choosing the final
unitary $C_l$ of a sequence given above is more constrained than
necessary for ensuring that the measurement outcome is deterministic
in the absence of error.  This may result in less effective error
depolarization.  We suggest two alternatives that greatly reduce the
constraints on $C_l$. The first is to choose $C_l$ uniformly at random
from all Clifford unitaries that ensure that the final state in the
absence of error is a logical state. This is equivalent to
constructing $C_l$ as an implementation of the inverse of the previous
unitaries followed by a random gate that can be decomposed into CNOT
and Pauli product operators.  An even more randomizing approach is to
choose $C_l$ as suggested in Ref.~\cite{Knill08}. In this case, $C_l$
is composed of a uniformly random Clifford unitary followed by
one-qubit Clifford gates randomly chosen to ensure that a randomly
chosen joint $Z$-measurement is deterministic in the absence of
errors. By a joint $Z$-measurement we mean measurement of a product of
$\sigma_z$ operators on a subset of the qubits. The product's
eigenvalues are $\pm 1$ and can be determined by multiplying the
standard basis measurement outcomes of the qubits in the subset, where
a qubit's $0$ and $1$ measurement outcomes are mapped to $1$, $-1$,
respectively.  This strategy takes advantage of the often much better
fidelity of one-qubit gates for the sequence-dependent part of the
last unitary.  A disadvantage is that instead of $n$ deterministic bit
values, only one bit value is obtained in each run of the
experiment. To fit the resulting data, we use Eqs.~\ref{Eq1n}
and~\ref{Eq4n} with $n=1$.

\section{Conclusion}
\label{sec:conclusion}

In summary, we have described a protocol for randomized benchmarking
of gates in a quantum information processor and implemented the
protocol experimentally on two qubits to measure the error per
operation of arbitrary two-qubit Clifford unitaries.  The protocol we
propose is independent of the gate set that is experimentally
implemented and so can provide an easily portable method for
evaluating the performance of Clifford unitaries on different physical
platforms. Furthermore, with this method it is straightforward to
isolate the fidelity of a specific two-qubit gate.  We have emphasized
some of the consistency checks that can be performed to qualify the
reported errors per operation or gate.  Looking ahead, this randomized
benchmarking protocol should prove useful as different experimental
implementations of quantum information processors aim to increase the
number of qubits and work to decrease the errors towards what is
required for fault-tolerance.

\begin{acknowledgments}
This work was supported by NSA, IARPA, ONR and the NIST Quantum
Information Program. The authors would like to thank Scott Glancy and
Brian Sawyer for careful readings of the manuscript.  JPG is supported
by NIST through an NRC fellowship.  This paper is a contribution by
NIST and not subject to U.S. copyright.
\end{acknowledgments}

\end{document}